\documentclass[letter]{ptptex}
\usepackage[dvips]{graphicx,color}
\usepackage{multirow}
\usepackage{bm}
\usepackage{wrapft}

\def\la{\mathrel{\mathpalette\fun <}}

\def\fun#1#2{\lower3.6pt\vbox{\baselineskip0pt\lineskip.9pt
\ialign{$\mathsurround=0pt#1\hfil##\hfil$\crcr#2\crcr\sim\crcr}}}

\title{
Dissociation of Relativistic Projectiles with the Continuum-Discretized
Coupled-Channels  Method
}

\author{
Kazuyuki \textsc{Ogata}$^1$ and
Carlos A. \textsc{Bertulani}$^2$
}

\inst{
$^1$Department of Physics, Kyushu University, Fukuoka 812-8581, Japan\\
$^2$Department of Physics, Texas A\&M University, Commerce, TX 75429, USA
}

\abst{
Relativistic effects in the breakup of weakly-bound nuclei at intermediate
energies are studied by means of the continuum-discretized coupled-channels
method with eikonal approximation.
Nuclear coupling potentials with Lorentz contraction are newly included
and those effects on breakup cross sections are investigated.
We show that relativistic corrections lead to larger breakup cross sections.
Coupled-channel effects on the breakup cross sections are also discussed.
}

\begin{document}
\maketitle

Reactions with radioactive nuclear beams are a major research area
in nuclear physics. The dissociation of weakly bound nuclei, or halo
nuclei, is dominated by the Coulomb interaction, although the
nuclear interaction with the target cannot be neglected in most
cases \cite{Isao1}. The final state interaction of the fragments
with the target, and between the themselves, leads to important
continuum-continuum and continuum-bound-state couplings which
appreciably modify the reaction dynamics. Higher-order couplings are
more relevant in the dissociation of halo nuclei due to their low
binding \cite{Ogata1,Ogata2}.

The continuum-discretized coupled-channels method (CDCC)
\cite{CDCC-review} is one of the most accurate models to describe
the breakup of halo nuclei taking account of higher-order couplings
explicitly. The eikonal CDCC method (E-CDCC) \cite{Ogata1,Ogata2},
which was developed by the Kyushu group, is a derivation of CDCC
that enables one to efficiently treat the nuclear and Coulomb
breakup reactions at $E_{lab} \ge 50$ MeV/nucleon. An essential
prescription described in Refs.~\citen{Ogata1,Ogata2} is the
construction of hybrid (quantum and eikonal) scattering amplitudes,
with which one can make quantum-mechanical (QM) corrections to the
pure eikonal wavefunctions with a minimum task. These corrections
are, however, expected to become less important as the incident
energy increases.

The eikonal CDCC equations are Lorentz covariant in the high energy
limit as shown later. However, this is only true if the Coulomb and
nuclear potentials used in the calculations are correspondingly
Lorentz covariant. This has not been explored, except for the
calculation presented in Ref.~\citen{ber05}. In fact, most rare
isotope facilities use projectile dissociation at 100--250
MeV/nucleon. At these energies, relativistic contraction of fields
and retardation effects \cite{ber03,ber07,hei07,esb08} are of the
order of 10--30 \%. Relativistic effects enter in the dynamics of
coupled-channels equations in a nonlinear, often unpredictable way,
which can lead to a magnification, or reduction,
of the corrections. In the present
work, we confirm the relevance of  the relativistic effects
mentioned above, henceforth called dynamical relativistic effects,
on the breakup cross sections of $^8$B and $^{11}$Be nuclei by
$^{208}$Pb target at 100 and 250 MeV/nucleon. We make use of E-CDCC
incorporating relativistic Coulomb and nuclear coupling potentials.
The role of the latter, a novel effect included in this work, is
investigated. We also see how the channel-coupling affects the
breakup cross section with and without dynamical relativistic
effects.

The multipole-expansion of the relativistic Coulomb potential
between the target nucleus (T) with the atomic number $Z_{\rm T}$
and the projectile (P), consisting of C and v clusters, are given in
Ref.~\citen{ber05}:
\begin{equation}
V_{{\rm E1}\mu}=\sqrt{\frac{2\pi}{3}}\xi Y_{1\mu}\left(  \mathbf{\hat
{\mbox{\boldmath$\xi$}}}\right)  \frac{\gamma Z_{\rm T}ee_{\rm E1}}{\left(
b^{2}+\gamma ^{2}z^{2}\right)^{3/2}}\left\{
\begin{array}
[c]{c}%
\mp b\ \ (\mathrm{if}\ \ \ \mu=\pm1)\\
\sqrt{2}z\ \ (\mathrm{if}\ \ \ \mu=0)\
\end{array}
\right.  \label{eq6}%
\end{equation}
for the E1 (electric dipole) field and%
\begin{align}
V_{{\rm E2}\mu}= &  \sqrt{\frac{3\pi}{10}}\xi^{2}Y_{2\mu}\left(
\mathbf{\hat {\mbox{\boldmath$\xi$}}}\right)  \frac{\gamma
Z_{\rm T}ee_{\rm E2}}{\left(  b^{2}+\gamma
^{2}z^{2}\right)  ^{5/2}}\nonumber\\
&  \times\left\{
\begin{array}
[c]{c}%
b^{2}\ \ \ \ (\mathrm{if}\ \ \ \mu=\pm2)\\
\mp(\gamma^2+1)bz\ \ \ \ (\mathrm{if}\ \ \ \mu=\pm1)\\
\sqrt{2/3}\left(  2\gamma^{2}z^{2}-b^{2}\right)  \ \ \ \ (\mathrm{if}%
\ \ \ \mu=0)\
\end{array}
\right.  \label{eq7}%
\end{align}
for the E2 (electric quadrupole) field. In
Eqs.~(\ref{eq6})--(\ref{eq7}), $e_{{\rm E}\lambda}=[Z_{\rm v}(A_{\rm
C}/A_{\rm P})^\lambda +Z_{\rm C}(-A_{\rm v}/A_{\rm P})^\lambda]e$
are effective charges for $\lambda=1$ and 2 multipolarities for the
breakup of ${\rm P}\rightarrow{\rm C}+{\rm v}$. The intrinsic
coordinate of v with respect to C is denoted by
$\mathbf{\mbox{\boldmath$\xi$}}$ and $b$ is the impact parameter (or
transverse coordinate) in the collision of P and T, which is defined
by $b=\sqrt{x^2+y^2}$ with ${\bf R} = (x,y,z)$, the relative
coordinate of P from T in the Cartesian representation. The Lorentz
contraction factor is denoted by $\gamma=\left(
1-v^{2}/c^{2}\right)^{-1/2}$, where $v$ is the velocity of P. Note
that these relations are obtained with so-called far-field
approximation \cite{EB05}, i.e. $R$ is assumed to be always larger
than $\xi$. The Coulomb coupling potentials in E-CDCC are obtained
with Eqs.~(\ref{eq6})--(\ref{eq7}) as shown below.

As for the relativistic nuclear potentials, we follow the conjecture
of Feshbach and Zabek \cite{FZ77}, in which Lorentz contraction was
introduced in a nuclear potential inspired on the folding model. In
the present work, we further make zero-range approximation to the
folding model. Accordingly, we replace the non-relativistic optical
potential $U({\bf b},z)$ between T and each constituent of P by
$\gamma U({\bf b},\gamma z)$. Even though this is a quite rough
prescription to include the dynamical relativistic corrections on
the nuclear potential, we can check effects of the correction on
breakup cross sections at least semi-quantitatively.

The E-CDCC equations for the three-body reaction under consideration
are given by \cite{Ogata1,Ogata2}:
\begin{equation}
\dfrac{i\hbar^2}{E_c}K_c^{(b)}(z)
\dfrac{d}{d z}\psi_{c}^{(b)}(z) = \sum_{c'}
{\mathfrak{F}}^{(b)}_{cc'}(z) \; {\cal R}^{(b)}_{cc'}(z) \;
\psi_{c'}^{(b)}(z) \ e^{i\left(K_{c'}-K_c \right) z}, \label{cceq4}
\end{equation}
where $c$ denotes the channel indices \{$i$, $\ell$, $m$\}; $i>0$
($i=0$) stands for the $i$th discretized-continuum (ground) state
and $\ell$ and $m$\ are respectively the orbital angular momentum
between the constituents (C and v) of the projectile and its
projection on the $z$-axis taken to be parallel to the incident
beam. Note that we neglect the internal spins of C and v for
simplicity. The impact parameter $b$ is relegated to a superscript
since it is not  a dynamical variable. The total energy and the
asymptotic wave number of P are denoted by $E_c$ and $K_c$,
respectively, and ${\cal R}_{cc'}^{(b)}(z)=(K_{c'} R-K_{c'}
z)^{i\eta_{c'}}/ (K_c R-K_c z)^{i\eta_c}$ with $\eta_c$ the
Sommerfeld parameter. The local wave number $K_c^{(b)}(z)$ of P is
defined by energy conservation as
\begin{equation}
E_c
=
\sqrt{(m_{\rm P}c^2)^2+\left\{\hbar c K_c^{(b)}(z)\right\}^2}
+
\dfrac{Z_{\rm P}Z_{\rm T}e^2}{R},
\label{klcl}
\end{equation}
where $m_{\rm P}$ is the mass of P and $Z_{\rm P}e$ ($Z_{\rm T}e$)
is the charge of P (T).
The reduced coupling-potential ${\mathfrak{F}}^{(b)}_{cc'}(z)$ is given by
\begin{equation}
{\mathfrak{F}}^{(b)}_{cc'}(z)
=
{\cal F}^{(b)}_{cc'}(z)
-\dfrac{Z_{\rm P}Z_{\rm T}e^2}{R}\delta_{cc'},
\label{FF1}
\end{equation}
where
\begin{equation}
{\cal F}^{(b)}_{cc'}(z)
=
\left\langle
\Phi_{c}
|
U_{\rm CT}+U_{\rm vT}
|
\Phi_{c'}
\right\rangle_{\bf \xi}
e^{-i(m'-m) \phi_R}.
\label{FF2}
\end{equation}
The $\Phi$ denotes the internal wavefunctions of P, $\phi_R$ is the
azimuthal angle of ${\bf b}$ and $U_{\rm CT}$ ($U_{\rm vT}$) is the
potential between C (v) and T consisting of nuclear and Coulomb
parts. In actual calculations we use the multipole expansion ${\cal
F}^{(b)}_{cc'}(z)=\sum_\lambda {\cal F}^{\lambda (b)}_{cc'}(z)$, the
explicit form of which is shown in Ref.~\citen{Ogata2}.

In order to include the dynamical relativistic effects described above,
we make the replacement
\begin{equation}
{\cal F}^{\lambda (b)}_{cc'}(z) \rightarrow \gamma f_{\lambda,m-m'}
{\cal F}^{\lambda (b)}_{cc'}(\gamma z) .\label{FF3}
\end{equation}
The factor $f_{\lambda,\mu}$ is set to unity for nuclear couplings,
while for Coulomb couplings we take
\begin{equation}
f_{\lambda,\mu}
=
\left\{
\begin{array}{cl}
1/\gamma & \quad (\lambda=1, \mu=0) \\
(\gamma^2+1)/(2\gamma)   & \quad (\lambda=2, \mu=\pm1) \\
1        & \quad ({\rm otherwise}) \\
\end{array}
\right.
\label{FF4}
\end{equation}
following Eqs.~(\ref{eq6}) and (\ref{eq7}). Correspondingly, we use
\begin{equation}
\dfrac{Z_{\rm P}Z_{\rm T}e^2}{R}\delta_{cc'}
\rightarrow
\gamma\dfrac{Z_{\rm P}Z_{\rm T}e^2}{\sqrt{b^2+(\gamma z)^2}}\delta_{cc'}
\label{FF5}
\end{equation}
in Eqs.~(\ref{klcl}) and (\ref{FF1}). The Lorentz contraction factor
$\gamma$ may have channel-dependence, i.e., $\gamma=E_c/(m_{\rm
P}c^2)$, which we approximate by the value in the incident channel,
i.e., $E_0/(m_{\rm P}c^2)$.

It should be remarked that we neglect the recoil motion of T in
Eq.~(\ref{cceq4}); this can be justified because we consider
reactions in which T is significantly heavier than P and we only
treat forward-angle scattering in the present study \cite{ber05}, as
shown below. Note also that in the high incident-energy limit ${\cal
R}_{cc'}^{(b)}(z)\to 1$ and $K_c^{(b)}(z) \to K_c$, unless the
energy transfer is extremely large. Thus, in this limit
Eq.~(\ref{cceq4}) becomes Lorentz covariant, as desired.

Using Eqs.~(\ref{cceq4})--(\ref{FF5}), we calculate the dissociation
observables in reactions of loosely bound nuclei $^8$B and $^{11}$Be
on $^{208}$Pb targets. The internal Hamiltonian of P and the number
of the states included are the same as in Ref.~\citen{Hussein}
except that we neglect the spin of the proton as mentioned above and
thus change the depth of the $p$-$^7$Be potential to reproduce the
proton separation energy of 137 keV. The optical potentials between
the constituents of P and T are the same as in Table I of
Ref.~\citen{Hussein}. Note that the results shown below do not
depend on the choice of these potentials significantly. The maximum
value of the internal coordinate $\xi$ is taken to be 200 fm. The
maximum impact parameter is set to be 500 fm and 450 fm for,
respectively, $^8$B and $^{11}$Be breakup reactions at 100
MeV/nucleon, while it is set to 400 fm for both reactions at 250
MeV/nucleon.

%
\begin{figure}[htpb]
\begin{center}
\includegraphics[width=0.55\textwidth,clip]{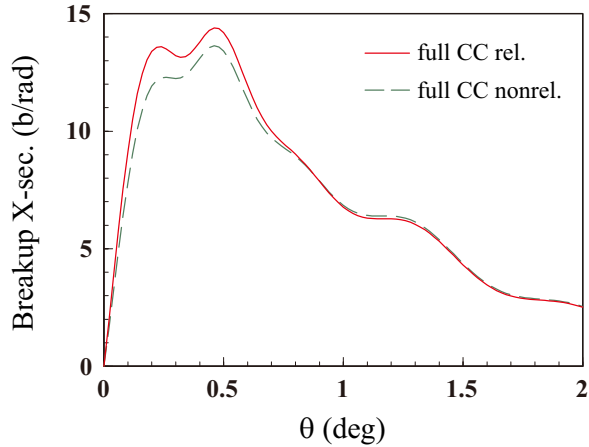}
\end{center}
\caption{
The total breakup cross section for $^8$B$+^{208}$Pb at
250 MeV/nucleon, as a function of the scattering angle
of the c.m. of the projectile after breakup.
The solid and dashed lines show the results of the full CC
calculation with and without the dynamical relativistic effects,
respectively.
}
\end{figure}
Figure 1 shows the total breakup cross section of $^8$B by
$^{208}$Pb at 250 MeV/nucleon, as a function of the scattering angle
$\theta$ of the center-of-mass (c.m.) of the projectile after
breakup. The solid and dashed lines represent the results of the
E-CDCC calculation with and without the dynamical relativistic
effects, respectively; in the latter we set $\gamma=1$ instead of
the proper value 1.268 in Eqs.~(\ref{FF3})--(\ref{FF5}). Note that
in all calculations shown in this work we use relativistic
kinematics, so that our results probe only the relativistic effects
on the dynamics.
One sees that the dynamical relativistic correction
gives significantly larger breakup cross sections for $\theta \la
0.7$ degrees; the difference between the two around the peak is
sizable, i.e. of the order of 10--15 \%.

Figure 2 shows the corresponding partial breakup cross sections as
a function of $b$.
%
\begin{figure}[htpb]
\begin{center}
\includegraphics[width=0.55\textwidth,clip]{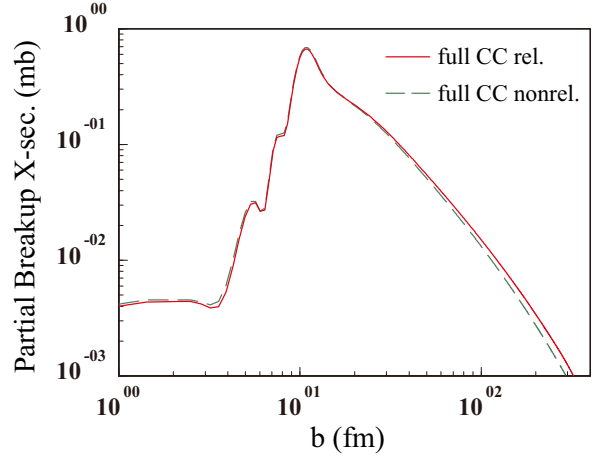}
\caption{
Partial breakup cross sections as a function of $b$
for $^8$B$+^{208}$Pb at 250 MeV/nucleon.}
\end{center}
\end{figure}
One sees that for $b \le 50$ fm the difference between the two is
negligibly small, while for $b > 50$ fm a clear enhancement of the
cross section due to dynamical relativistic effects is found. Since
the nuclear coupling potentials in E-CDCC calculations for the
reaction under study are limited to $b$ less than about 15 fm, at
the most, the enhancement of the breakup cross section shown in
Fig.~1 is due to the dynamical relativistic correction to the
Coulomb potential. In other words, the effects of relativistic
corrections in the nuclear potentials are negligible, which is a new
important finding in the present study. This can be seen more
clearly in Fig.~3, for the breakup cross sections calculated with
the E-CDCC method, with only the nuclear coupling potentials. The
relativistic and non-relativistic results in Fig.~3 agree very well
with each other.
%
\begin{figure}[htpb]
\begin{center}
\includegraphics[width=0.55\textwidth,clip]{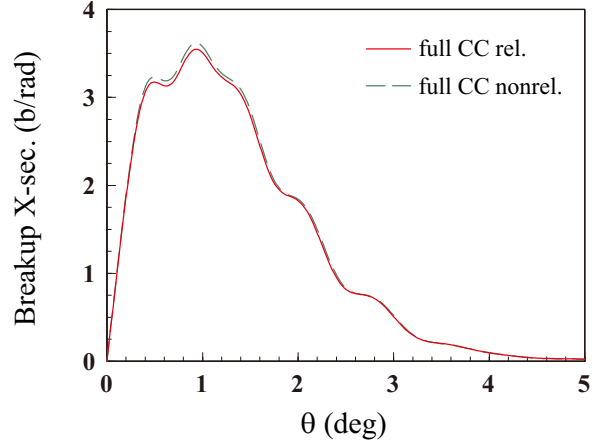}
\caption{ Same as in Fig.~1 but only the nuclear coupling potentials
are included in the E-CDCC calculation. }
\end{center}
\end{figure}

Next we investigate how the coupled-channel calculations affect the
breakup cross section and the role of dynamical relativistic
corrections. For this purpose, a first-order perturbative calculation
is performed. This first-order calculation is consistent with the
equivalent photon method, as described in Ref.~\citen{ber88}.
In fact, firs-order Coulomb excitation can be expressed as
$d\sigma/dE_\gamma=N_{E\lambda}(E_\gamma)
\sigma_\gamma^{(E\lambda)}(E_\gamma)$,
where $N_{E\lambda}(E_\gamma)$ is the equivalent photon spectrum
for the $E\lambda$ multipolarity, and
$\sigma_\gamma^{(E\lambda)} (E_\gamma)$ is the
corresponding photonuclear dissociation cross section.
Using the expressions for $N(E\lambda)$, $\lambda=1$, 2,
given in Ref.~\citen{ber88} with
the matrix elements for the $E\lambda$ operator used in the present work,
we confirm that first order perturbation theory and the
equivalent photon method yield exactly the same results, as expected.

We show in Fig.~4 the results of full-CDCC and the first-order
calculation; the left (right) panel corresponds to the calculation
with both nuclear and Coulomb breakup (only Coulomb breakup).
%
\begin{figure}[htpb]
\begin{center}
\includegraphics[width=1.0\textwidth,clip]{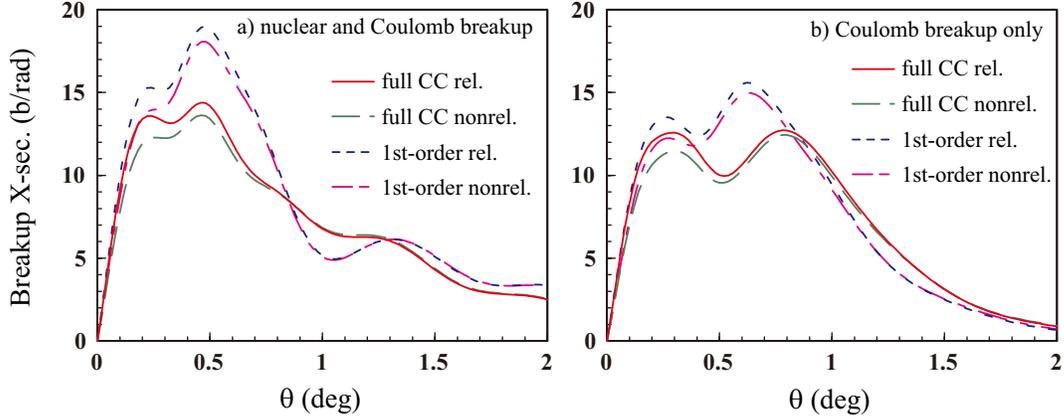}
\caption{
The total breakup cross sections for $^8$B$+^{208}$Pb at
250 MeV/nucleon with nuclear and Coulomb breakup (left panel)
and only Coulomb breakup (right panel).
The solid (dotted) and dashed (dash-dotted) lines show
the results of the full CC (first-order perturbative)
calculation with and without the relativistic correction, respectively.
}
\end{center}
\end{figure}
In each panel the solid (dotted) and dashed (dash-dotted) lines show
the results of the full CC (first-order perturbative) calculation
with and without the dynamical relativistic correction,
respectively.
%
\begin{figure}[b]
\begin{center}
\includegraphics[width=0.55\textwidth,clip]{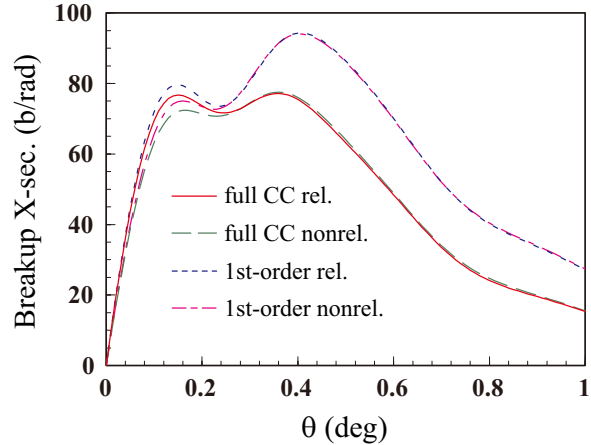}
\caption{
Same as in Fig.~4a but for $^{11}$Be$+^{208}$Pb at 250 MeV/nucleon.
}
\end{center}
\end{figure}
One sees that relativistic corrections modify the first-order results
in the same way as they do with the full CC calculation.
We stress here, however, that since continuum-continuum
couplings make relativistic effects non-linear (and non-trivial to
interpret), one cannot infer the effect of relativistic
corrections by simply carrying out first-order calculations.
More seriously, the full CC and first-order calculations give quite
different breakup cross sections even at forward angles.
Full CC calculation is necessary to obtain
a reliable breakup cross section to be compared with experimental
data. In other words, continuum-continuum couplings are important
in describing breakup processes even at intermediate energies.
It is found that continuum-continuum couplings for both
nuclear and Coulomb parts play significant roles.

In Fig.~5 we show the results for $^{11}$Be breakup by $^{208}$Pb at
250 MeV/nucleon, with $\gamma=1.268$. Differences between the
relativistic and nonrelativistic calculations appear below about 0.3
degrees for both full CC and first-order
perturbative results, and the increase of the cross section
around the peak is,
as for the $^{8}$B breakup, about 10--15 \%.

Figures 6 and 7 show, respectively, the results for $^8$B$+^{208}$Pb
and $^{11}$Be$+^{208}$Pb at 100 MeV/nucleon. The main features of
the results are the same as at 250 MeV/nucleon, except that the
effects of relativity are somewhat reduced, i.e., the enhancement of
the cross section at the peak is below the 10\% level.
%
\begin{figure}[htpb]
\begin{center}
\includegraphics[width=0.55\textwidth,clip]{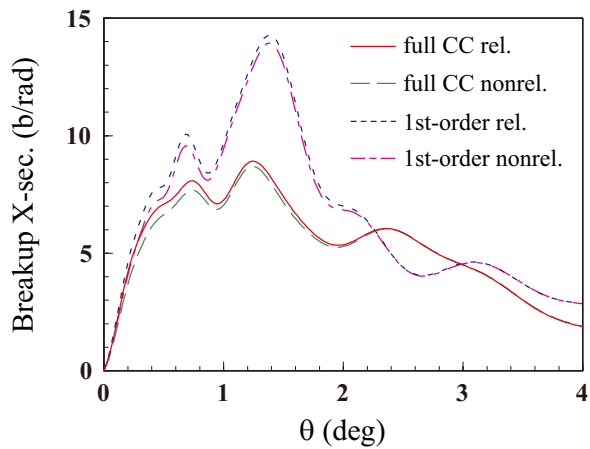}
\caption{
Same as in Fig.~4a but for $^{8}$B$+^{208}$Pb at 100 MeV/nucleon.
}
\end{center}
\end{figure}
%
%
\begin{figure}[htpb]
\begin{center}
\includegraphics[width=0.55\textwidth,clip]{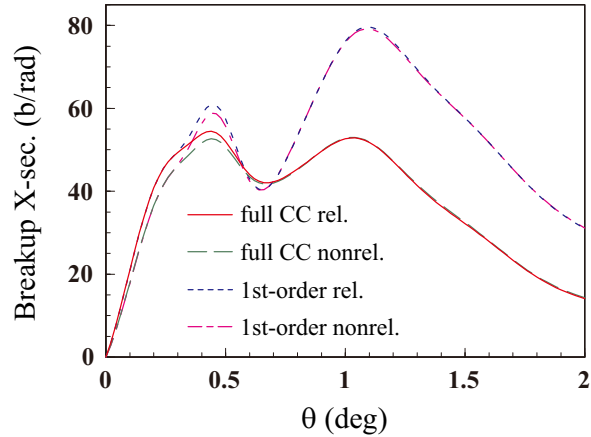}
\caption{
Same as in Fig.~4a but for $^{11}$Be$+^{208}$Pb at 100 MeV/nucleon.
}
\end{center}
\end{figure}
This rather small difference can still be important for some
quantitative analysis, e.g. determination of the astrophysical
factor $S_{17}$ for the $^7$Be($p,\gamma$)$^8$B reaction through
$^8$B breakup reaction. In order to draw a definite conclusion,
however, we need to quantitatively examine the approximations used
to derive Eqs.~(\ref{eq6}) and (\ref{eq7}), i.e., use of point
charge for C, v and T, and also the far-field approximation
\cite{EB05}. Moreover,  an evaluation of quantum mechanical
corrections to the breakup cross sections, which can be done by
constructing hybrid scattering amplitudes \cite{Ogata1,Ogata2}, will
be necessary. Nevertheless, relativistic effects on the breakup cross
sections of about 15\% found at 250 MeV/nucleon needs to be
seriously addressed in the future.

In conclusion, we have evaluated the effects of relativistic
corrections of the nuclear and Coulomb coupling potentials on the
breakup cross sections of the weakly bound projectiles $^8$B and
$^{11}$Be by $^{208}$Pb targets at 250 and 100 MeV/nucleon. The
relativistic corrections modify appreciably the breakup cross
sections, at the level of 15\% (10\%), in collisions at  250 (100)
MeV/nucleon. This change is found to be due mainly to the
modification of the Coulomb potential. We have shown that
continuum-continuum couplings are also influenced by
relativistic corrections and modify breakup cross sections appreciably.
These important features have been widely ignored in the literature
and deserve further theoretical studies. We have found quite strong
relativistic effects on breakup energy spectra of $^8$B. More
detailed and systematic analysis including this subject will be
presented in a forthcoming paper.

This work was partially supported  by the U.S. DOE grants
DE-FG02-08ER41533 and DE-FC02-07ER41457 (UNEDF, SciDAC-2), and the JUSTIPEN/DOE
grant DEFG02- 06ER41407.
The computation was carried out using the computer facilities at
Research Institute for Information Technology, Kyushu University.


\begin{thebibliography}{99}

\bibitem{Isao1}
I. Tanihata, Prog. Part. Nucl. Phys. {\bf 35}, 505 (1995) and
references cited therein.

\bibitem{Ogata1}
K. Ogata, M. Yahiro, Y. Iseri, T. Matsumoto and M. Kamimura,
Phys. Rev. C {\bf 68}, 064609 (2003).

\bibitem{Ogata2}
K. Ogata, S. Hashimoto, Y. Iseri, M. Kamimura and M. Yahiro,
Phys. Rev. C {\bf 73}, 024605 (2006).

\bibitem{CDCC-review}
M. Kamimura {\it et al}., Prog. Theor. Phys. Suppl. {\bf 89}, 1 (1986);
N. Austern {\it et al}., Phys. Rep. {\bf 154}, 125 (1987).

\bibitem{ber05}C. A. Bertulani, Phys. Rev. Lett. {\bf 94}, 072701 (2005).

\bibitem{ber03}
C. A. Bertulani, A. E. Stuchbery, T. J. Mertzimekis and A. D. Davies,
Phys. Rev. C {\bf 68}, 044609 (2003).

\bibitem{ber07}
C. A. Bertulani, G. Cardella, M. De Napoli, G. Raciti, E. Rapisarda,
Phys. Lett. {\bf B650}, 233 (2007).

\bibitem{hei07}
H. Scheit, A. Gade, Th. Glasmacher, T. Motobayashi, Phys. Lett. {\bf
B659}, 515 (2007).

\bibitem{esb08}
H. Esbensen, Phys. Rev. {\bf C78}, 024608 (2008).

\bibitem{EB05} H. Esbensen and C.A. Bertulani, Phys. Rev. C {\bf 65}, 024605 (2002).

\bibitem {FZ77}H. Feshbach and M. Zabek, Ann. of Phys. {\bf 107} (1977) 110.

\bibitem{Hussein}
M.~S. Hussein, R. Lichtenth\"{a}ler, F.~M. Nunes and I.~J. Thompson,
Phys. Lett. {\bf B640}, 91 (2006).

\bibitem{ber88}
C. A. Bertulani and G. Baur, Phys. Rep. {\bf 163}, 299 (1988).

\end{thebibliography}
\end{document}